\documentclass[reprint,amsmath,amssymb,prl,superscriptaddress]{revtex4-2}

\usepackage{physics}
\usepackage{braket}
\usepackage{graphicx}
\usepackage{xcolor}
\usepackage{hyperref}
\usepackage[capitalise]{cleveref}
\usepackage{placeins}

\newcommand{\pol}{\boldsymbol{\epsilon}}
\newcommand{\Pt}{\mathbf{P}_\text{tot}}
\newcommand{\Qph}{Q_\text{ph}}

\newcommand{\Piph}{\Pi_\text{ph}}
\newcommand{\Vext}{v_\text{ext}}

\newcommand{\Ec}{E_\text{c,\,el-ph}}

\begin{document}
\title{The light-matter correlation energy functional of the cavity-coupled two-dimensional electron gas via quantum Monte Carlo simulations}
\author{Lukas Weber}
\affiliation{Center for Computational Quantum Physics, The Flatiron Institute, 162 Fifth Avenue, New York, NY 10010, USA}
\affiliation{Max Planck Institute for the Structure and Dynamics of Matter, Luruper Chaussee 149, 22761 Hamburg, Germany}
\author{Miguel A. Morales}
\affiliation{Center for Computational Quantum Physics, The Flatiron Institute, 162 Fifth Avenue, New York, NY 10010, USA}
\author{Johannes Flick}
\affiliation{Center for Computational Quantum Physics, The Flatiron Institute, 162 Fifth Avenue, New York, NY 10010, USA}
\affiliation{Department of Physics, City College of New York, New York, New York 10031, United States}
\affiliation{Physics Program, Graduate Center, City University of New York, New York New York 10016, United States}
\author{Shiwei Zhang}
\affiliation{Center for Computational Quantum Physics, The Flatiron Institute, 162 Fifth Avenue, New York, NY 10010, USA}
\author{Angel Rubio}
\affiliation{Max Planck Institute for the Structure and Dynamics of Matter, Luruper Chaussee 149, 22761 Hamburg, Germany}
\affiliation{Center for Computational Quantum Physics, The Flatiron Institute, 162 Fifth Avenue, New York, NY 10010, USA}

\begin{abstract}
We
perform extensive
simulations of the two-dimensional cavity-coupled electron gas in a modulating potential as a minimal model for cavity quantum materials. 
These simulations are enabled by a newly developed quantum-electrodynamical (QED) auxiliary-field quantum Monte Carlo method.
We present a procedure to greatly reduce finite-size effects in such calculations.
Based on our results, we show
that a modified version of weak-coupling perturbation theory is remarkably accurate for a large parameter region. We further provide a simple parameterization of the light-matter correlation energy as a functional of the cavity parameters and the electronic density. These results provide a numerical foundation for the development of the QED density functional theory, which was previously reliant on analytical approximations, 
to allow quantitative modeling of a wide range of systems with light-matter coupling.
\end{abstract}
\maketitle

In recent years, cavity quantum electrodynamics (QED)~\cite{ruggenthaler_quantumelectrodynamical_2014,Hubener_2020,schlawin_cavity_2022,Hubener2024,ruggenthaler_understanding_2022}  has started to unveil its potential as a tool to modify not only chemical reactions~\cite{thomas2019tilting,ahn2023modification,ebbesen2023} but also the properties of quantum matter~\cite{appugliese_breakdown_2022,jarc_cavitymediated_2023}. These advances have created a demand for predictive numerical methods that can treat cavity photons and the constituents of matter on the same level. To satisfy this demand, extensions of existing electronic-structure methods have been developed, such as QED coupled-cluster methods~\cite{haugland_coupled_2020,pavosevic_polaritonic_2021,pavosevic_computational_2023}, density-matrix renormalization group methods~\cite{eckhardt_quantum_2022,passetti_cavity_2023,shaffer_entanglement_2024}, and
quantum Monte Carlo (QMC) methods, including the stochastic series expansion~\cite{weber_cavityrenormalized_2023,langheld_quantum_2024}, and 
diffusion Monte Carlo~\cite{weight_diffusion_2024}.

While the majority of these algorithms emphasize accurate computations for molecular or lattice-model systems, there is currently no comprehensive many-body approach for cavity-coupled bulk systems, in particular in the continuum.
A promising candidate for ab-initio calculations for such systems is the quantum electrodynamical density functional theory (QEDFT)~\cite{ruggenthaler_quantumelectrodynamical_2014,tokatly_timedependent_2013,flick_kohnsham_2015}, which has seen tremendous progress in the development of light-matter exchange-correlation energy functionals. Progress in the development of these functionals has come from multiple approaches, including the optimized effective potential method~\cite{pellegrini_optimized_2015},   the fluctuation-dissipation theorem~\cite{flick_simple_2022}, the random-phase approximation~\cite{novokreschenov_quantum_2023}, the many-body-dispersion method~\cite{tasci_photon_2024}, or the photon-free approximation~\cite{schafer_making_2021,lu_electronphoton_2024}.
These functionals and their underlying approximations have been shown to work well in selected regimes of light-matter coupling or frequency. However, they still lack the foundation of generalizable and reliable numerical results covering the whole range of interactions, at a level comparable to the QMC solution of the electron gas~\cite{ceperley_ground_1980} that 
originally enabled the success of purely electronic DFT~\cite{perdew_selfinteraction_1981}.
 
\begin{figure}
\begin{center}
\hspace{1cm}\includegraphics{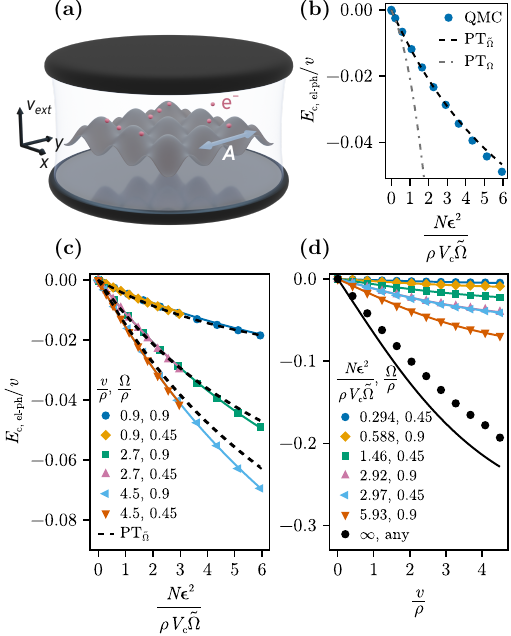}\\
\end{center}
\caption{\textbf{Electron-photon correlation energy} of the cavity-coupled electron gas in the modulated potential. (a) Sketch of the model with electron gas in a soft external potential $\Vext$. The polarization of the cavity mode is in one of the lattice directions. (b) The unreasonable effectiveness of weak-coupling perturbation theory with $\tilde{\Omega}=\text{const}$ (PT$_{\tilde{\Omega}}$) compared to leaving $\Omega=\text{const}$ (PT$_{\Omega}$) and QMC data for $v/\rho = 2.7$ and $\Omega/\rho=0.9$. (c-d) Fit (solid lines) of the light-matter correlation energy $\Ec$ to slices of the parameter space, spanned by the average electron density $\rho=1/a^2$, the potential depth $v$, the frequency $\Omega$, and the light matter coupling $|\pol|$. Black dashed lines correspond to the weak-coupling perturbation theory (PT$_{\tilde{\Omega}}$), while black circles are AFQMC results of the infinitely light-matter coupled asymptotic model, which was not included in the fit. The black solid line shows the extrapolation of the fit to infinite coupling. The errorbars are smaller than the symbol size.}
\label{fig:gas}
\end{figure}

In this paper, we provide this numerical foundation by applying the newly developed
QED auxiliary-field quantum Monte Carlo
(QED-AFQMC) method~\cite{weber_phaseless_2024} to solve a minimal model for cavity quantum materials: the cavity-coupled two-dimensional electron gas in a soft modulating external potential~(\cref{fig:gas}(a)). 
After proposing a scheme to handle a class of finite-size effects that uniquely plague QED-coupled bulk systems, we obtain the light-matter correlation energy in the thermodynamic limit.
For weak potentials, this energy is described well by a modified perturbation theory. For the full range of parameters, we fit the correlation energy to a simple functional of the cavity parameters and the electronic density. These findings serve as both a benchmark for future cavity QED many-body methods and as a direct ingredient for the ongoing QEDFT functional development.

\textit{Model}. We consider the Hamiltonian~\cite{cohen_photons_1997} of many  
electrons coupled to a single photon mode of frequency $\Omega$
\begin{align}
\label{eq:model}
H &= \sum_{i} \frac{(\mathbf{p}_i + \mathbf{A})^2}{2} + \Vext(\mathbf{r}_i) + \frac \Omega 2 (\Piph^2 + \Qph^2 - 1)
\end{align}
in two dimensions, where $\mathbf{A} = \Qph \pol/\sqrt{\Omega V_\text{c}}$, where $\Qph$ ($\Piph$) is the photon displacement (momentum) operator of the photon mode. We treat the model in the long-wavelength limit, with a position independent vectorial coupling $\pol$ and a mode volume $V_\text{c}$, which we scale with the system size. 
Our method, which can treat realistic electronic interactions in periodic systems in the continuum at high accuracy~\cite{weber_phaseless_2024,motta_initio_2018}, allows the incorporation of
light and electron-light interaction on equal footing.
As a first step, in this work, we will not consider Coulomb interactions. Under this condition, our AFQMC simulations do not suffer from a phase problem~\cite{zhang_quantum_2003} and are exact.

The cavity-coupled homogeneous electron gas ($\Vext = 0$), due to the conservation of the total momentum, can be solved exactly~\cite{rokaj_free_2022} by a product state between light and matter, $\ket{\Psi_\text{el}}\otimes \ket{\chi_\text{ph}}$, completely lacking any light-matter correlation. In a real material (with $\Vext \neq 0$), however, the crystal lattice breaks the perfect momentum conservation and induces momentum fluctuations that render the light-matter interaction nontrivial.

Therefore, for a true minimal model of cavity materials, we need to add an external potential to break the translation symmetry (\cref{fig:gas}(a)).
In this work, we choose a cosine potential $\Vext(\mathbf{r}) = -v\sum_{d=x,y} \cos(\frac{2\pi \mathbf{e}_d}{a} \cdot \mathbf{r})$
with the potential depth $v$ and the lattice constant $a$.
While this choice is primarily to keep the computational demand in our plane-wave-basis calculation manageable, it could in principle also be realized as a moiré potential~\cite{zhang_moire_2024}.

\begin{figure}
\begin{center}
\includegraphics{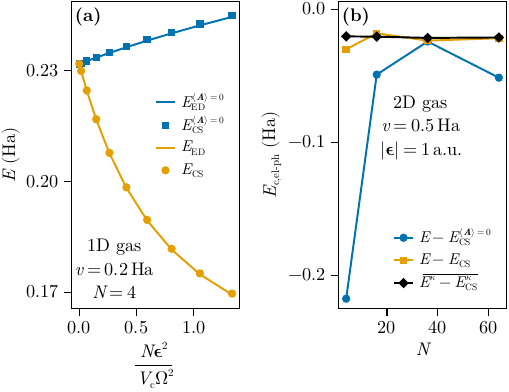}
\end{center}
\caption{\textbf{Toroidal magnetic finite-size effects.} In both panels, $a = 3\,a_0$, $\Omega = 0.05\,\text{Ha}$, and $N=N_\text{uc}$ is the number of electrons and unit cells. (a) The low-energy spectrum of the one-dimensional cavity-coupled electron gas computed with exact diagonalization. At finite light-matter coupling, the ground state, $E_\text{ED}$, is a doublet belonging to total crystal momentum $K = \pm 4\pi/L = \pm \pi/a$ (yellow circles) and correspondingly finite $\braket{\mathbf A}$. The first excited state, $E_\text{ED}^{\braket{\vb{A}}=0}$, is a doublet with $K=0$. The Hilbert space was constrained to 21 momenta and $n_\text{ph} < 20$, which is converged to the complete basis set limit. (b) A comparison of the finite-size scaling of the two-dimensional light-matter correlation energy derived from different subtraction schemes, obtained from AFQMC. $\overline{E^\kappa - E^\kappa_\text{CS}}$ denotes the average over 60 twisted boundary conditions.} 
\label{fig:scaling_comp}
\end{figure}

\textit{Toroidal magnetic finite-size effects}. When performing correlated calculations on finite systems, care has to be taken to correctly deal with finite-size effects. We find that light-matter coupled systems with periodic boundary conditions display a particular class of finite-size effect that---unless subtracted---dominates the effect of light-matter interactions on the energy in numerically accessible system sizes (\cref{fig:scaling_comp}(a)).

Considering our Hamiltonian from \cref{eq:model} with periodic boundary conditions, or equivalently, on a finite torus, the mode function of the cavity photon, albeit constant, wraps around the torus in one direction, realizing a finite magnetic flux loop. In finite systems with periodic boundary conditions, this can in general give rise to a symmetry broken ground state with $\braket{\vb A}\neq 0$ that describes a magnetic field induced by a circular current.
Another way to interpret this phenomenon is to consider that a finite momentum space lattice breaks the gauge invariance of the model that would usually allow eliminating any finite expectation value of $\braket{\vb A}$ via a gauge transformation $\vb p_i\mapsto \vb p_i - \braket{\vb A}$.
While these finite toroidal currents and fields have to vanish in the thermodynamic limit, they lead to strong energy contributions that can mask the signal of the light-matter correlation energy, making the proper extrapolation to the thermodynamic limit difficult (\cref{fig:scaling_comp}(b)).

We therefore propose a combination of two strategies to mitigate this problem. First, we define the light matter correlation energy as $\Ec = E - E_\text{CS}$ where $E_\text{CS}$ is the best variational energy that can be achieved with a light-matter product state on the same finite system, explicitly allowing $\braket{\vb A}\neq 0$. Both terms of this difference display magnetic solutions, leading to a leading-order cancellation of the term $\braket{\Pt}\cdot \braket{\mathbf{A}}$ from the energy (\cref{fig:scaling_comp}(b)).

In the absence of Coulomb interactions, we can obtain $E_\text{CS}$ by variationally minimizing the energy of a product state between a Slater-determinant and a squeezed coherent state (CS): $\ket{\Psi_\text{CS}} = \ket{\psi_\text{el}}\otimes\int dq \sqrt[4]{\frac s \pi} e^{-s (q-q_0)^2/2}\ket{q}$ with the parameters $\ket{\psi_\text{el}}$, $s$, and $q_0$. Under the assumption that the ground state is a pure light-matter product state, this ansatz solves \cref{eq:model} exactly, such as in the case $v=0$. We will further use this state as the trial wave function in our AFQMC calculations.

In addition to the introduction of this new
definition of the correlation energy, we also employ twist-averaged boundary conditions~\cite{lin_twistaveraged_2001}. These boundary conditions can be realized by quantizing the momenta in the $L\times L$ supercell as $\mathbf{k_n} = \frac{2\pi}{L} (\mathbf{n} + \boldsymbol{\kappa})$ with a shift $\boldsymbol{\kappa}$ that lives in the square $\Vert\boldsymbol\kappa\Vert_\infty < \frac 12$ and then averaging the results over $\boldsymbol\kappa$. On top of being an established approach to reduce QED-unrelated finite-size effects~\cite{qin_benchmark_2016}, the average over $\boldsymbol{\kappa}$ formally restores gauge invariance, so that e.g. the twist-averaged $\overline{\braket{\mathbf{A}}}$ or $\overline{\braket{\Pt}}$ vanish, and convergence with system size is vastly improved (\cref{fig:scaling_comp}(b)). With this approach, we find sufficient convergence to the thermodynamic limit already at $N=4\times 4$ unit cells. To perform the twist-average efficiently, we
use 
quasi-random 
$\boldsymbol\kappa$ 
~\cite{qin_benchmark_2016},
chosen from the low-discrepancy Halton sequence~\cite{halton_efficiency_1960}.

\textit{Results}.
In our calculations, we will consider supercells with one electron per $a\times a$ unit cell. We build a plane-wave basis consisting of the momenta $\mathbf{k_n}$ where we constrain the integer vector $\mathbf{n}$ to the smallest sphere that contains at least $f N_\text{uc}$ momenta and average over 60 twist angles. For our $4\times 4$ supercell we find that $f=15$ is sufficient to converge to the complete basis set limit. We perform our AFQMC calculations at a time step of $\tau=0.01\,\text{Ha}^{-1}$ with 300 walkers. 
We have verified
that the residual systematic error from either of these parameters is smaller than our statistical error bars.

To map out the parameter space, we first identify the different energy scales in our model, namely the kinetic energy of the electrons, $E_\text{kin} \sim \rho^{2/D}$ (in $D=2$ dimensions), with $\rho=1/a^2$ for one electron per unit cell, and the potential depth $v$, the renormalized cavity frequency $\tilde{\Omega} = \Omega \sqrt{1 + N\pol^2 /V_\text{c}\Omega^2}$, and the
light-matter interaction scale $N \pol^2/V_\text{c}\tilde{\Omega}$. Note that the results can only depend on $\Omega$ through $\tilde{\Omega}$ as can be shown by absorbing the $\mathbf{A}^2$ term in a canonical transformation.

Using these energy scales, we can construct dimensionless ratios and write the light-matter correlation energy as
\begin{equation}
\frac{\Ec}{\rho^{2/D}} = \mathcal{E}(x_v, x_\epsilon, x_{\tilde{\Omega}})
\end{equation}
with $x_v = \rho^{-2/D} v$, $x_\epsilon = \rho^{-2/D} N\pol^2/V_\text{c}\tilde{\Omega}$, and $x_{\tilde{\Omega}} =  \rho^{-2/D}\tilde{\Omega}$.

We perform simulations of slices through this parameter space, either along the direction of $x_\epsilon$ (\cref{fig:gas}(c)) or $x_v$ (\cref{fig:gas}(d)). For both $x_\epsilon = 0$ or $x_v = 0$, the correlation energy vanishes---the latter being consistent with the fact that for the homogeneous electron gas, the ground state is a light-matter product state. Conversely, for $x_\epsilon$ approaching infinity, the correlation energy approaches a bound given by an asymptotic infinite-coupling Hamiltonian, 
\begin{align}
H_\infty &= \lim_{|\pol|\to\infty} U_\text{LF}^\dag H U_\text{LF}\nonumber\\
&= \frac{1}{2} \sum_i \qty[ \mathbf p_i^2 +\Vext\qty(\mathbf r_i)] - \frac{(\hat{\pol}\cdot\Pt)^2}{2N},
\end{align}
with $\hat{\pol}=\pol/|\pol|$,
which follows from the Lang-Firsov transformation,
$U_\text{LF}=\exp(i\Piph \pol\cdot\Pt/\sqrt{V_\text{c} \tilde{\Omega}^3})$,
 as shown in the SM~\cite{SM}. This Hamiltonian can also be simulated exactly using a standard AFQMC calculation and its correlation energy is shown in \cref{fig:gas}(d). The original photon-free functional~\cite{schafer_making_2021} for QEDFT approximates the light-matter problem with a similar Hamiltonian. It has been shown to perform especially well in the strong-coupling regime. However, we find that convergence to the strong-coupling limit is in general slow, which explains why a functional based purely on a strong-coupling approximation does not generalize well to the weaker coupling regimes and can be improved by interpolating to a weak-coupling approximation~\cite{lu_electronphoton_2024}.

We further compare our data to weak-coupling perturbation theory (PT) for the light-matter correlation energy, derived in the SM~\cite{SM},
\begin{equation}
\label{eq:perturb}
\Ec^\text{PT$_{\tilde{\Omega}}$} = - \frac{1}{V_\text{c} \tilde{\Omega}} \sum_{\substack{n \in \text{occ}\\m\notin\text{occ}}} \frac{|\!\braket{\psi_m^\text{el}|\pol\cdot \vb p|\psi_n^\text{el}}\!|^2}{\tilde{\Omega} + \varepsilon_m^\text{el} - \varepsilon^\text{el}_n}.
\end{equation}
Here, $\ket{\psi^\text{el}_n}$ and $\varepsilon_n^\text{el}$ are the single-particle eigenstates and energies of the matter system, with the same twist-averaged boundary conditions as the other calculations, and $\vb p$ is the single-particle momentum. In our derivation we have left $\tilde{\Omega}$ intact, which we call PT$_{\tilde{\Omega}}$. Alternatively one could expand $\tilde{\Omega}=\Omega + \mathcal{O}(\pol^2)$, which we call PT$_{\Omega}$.

\cref{fig:gas}(b,c) shows that while PT$_{\Omega}$ is only valid for small light-matter coupling, PT$_{\tilde{\Omega}}$ is remarkably accurate to very large couplings, as long as $x_v$ is not too large. In earlier calculations in the velocity gauge, PT$_{\tilde{\Omega}}$ has been a natural choice~\cite{rivera_variational_2019,lu_electronphoton_2024}, but this choice is much less obvious in the dipole gauge. Due to the gauge invariance of the perturbation theory, it may be possible to simply replace $\Omega=\tilde{\Omega} \sqrt{1-N\pol^2/V_\text{c}\tilde{\Omega}^2}$ in the dipole gauge Hamiltonian to significantly enhance dipole-gauge perturbative calculations.

Our next goal is to find a simple analytical expression for $\mathcal{E}$ as a functional of the electron density and the cavity parameters that fits our numerical data well. To this end, we must first eliminate the explicit dependence on $x_v$. Since the photon only sees the spatially integrated effect of the momentum fluctuations caused by the potential, it is expected that the average of the squared gradient of the density $n(\vb r)$ in the polarization direction,
\begin{equation}
\mathcal{Q}^2 = \frac{\int dV (\hat{\pol}\cdot \nabla n(\vb r))^2}{\int dV (n(\vb r))^2}
\end{equation}
is a good proxy for the effect of $x_v$.

\begin{figure}
\begin{center}
\includegraphics{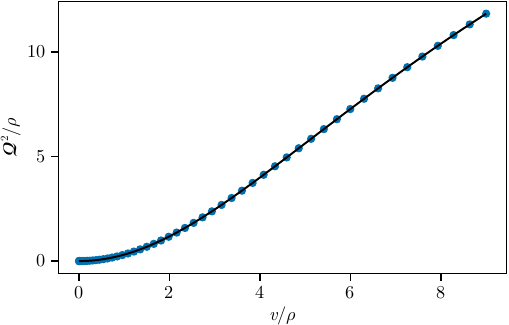}
\end{center}
\caption{Average density gradient fluctuations in the polarization direction, $\mathcal{Q}^2$, of the modulated electron gas, as a function of the squared potential depth $v^2$. Obtained from a noninteracting calculation. The solid line is a fit as described in the main text.}
\label{fig:gradient_mapping}
\end{figure}
Consequently, we want to determine the dimensionless function $x_q(x_v) = \mathcal{Q}^2/\rho^{2/D}$. Since in our extended system, electronic quantities are not affected by the intensive contribution of the photons, we can calculate this quantity in the pure matter system (\cref{fig:gradient_mapping}). To this data, we fit the function $x_q(x_v) = x_v^2/(b_1 + b_2 x_v + b_3 x_v^{3/2})$, where the exponent $\frac{3}{2}$ can be explained by the asymptotic behavior for $v\to \infty$~\cite{SM}. We find the parameters $b_1=3.27$, $b_2=-0.242$, and $b_3 = 0.213$.

Having found a proxy for the $x_v$ dependence, the next step is to come up with a fitting function $\mathcal{E}(x_q, x_\epsilon, x_{\tilde{\Omega}})$. A first helpful observation is that the parameter $x_{\tilde{\Omega}}$ barely contributes at fixed $x_v$ and $x_\epsilon$ (\cref{fig:gas}(c)). This is not surprising since $x_{\tilde{\Omega}} \stackrel{x_\epsilon \to \infty}{\longrightarrow} x_\epsilon$. Additionally, we can take advantage of the asymptotics of the perturbative solutions for weak coupling, $\mathcal{E}\stackrel{x_\epsilon\to 0}{\sim} x_\epsilon x_q$, and strong coupling, $\mathcal{E}\stackrel{x_\epsilon\to\infty}{\sim} \mathcal{E}(x_q)$. 
We propose a simple rational function $\mathcal{E}(x_v,x_\epsilon) = -\frac{x_\epsilon x_q}{c_1 x_\epsilon + c_2}$
that fulfills these criteria and has two parameters we fit to the data from \cref{fig:gas}(c--d). While it is possible to also include the $|\pol|=\infty$ data in the fit, for the sake of simplicity of the functional and accuracy at realistic couplings, we only fit to finite coupling, where our maximum deviation from the data is 0.4 mHa. Nevertheless, our functional still extrapolates qualitatively correctly to the infinite coupling case (\cref{fig:gas}(d)).

Reexpressing in the dimensionful quantities, we arrive at the functional
\begin{equation}
	\label{eq:params}
\Ec \approx -\frac{\mathcal{Q}^2}{c_1 + c_2 \rho^{\frac 2D} \frac{V_\text{c} \tilde{\Omega}}{N \pol^2}}\,,
\end{equation}
with the fit parameters $c_1=4.672$, $c_2=63.73$.

\textit{Conclusion.} We have performed high-accuracy AFQMC simulations for the cavity-coupled two-dimensional electron gas in a modulating external potential. 
As a first step, we study the system without Coulomb interactions and with a single photon mode in the long-wavelength approximation. After dealing with a class of magnetic finite-size effects that uniquely affect cavity-coupled systems on finite tori, we extract the light-matter correlation energy, scanning the model parameter space including the full range of light-matter couplings. Finally we provide a parametrization of this energy, in terms of the light-matter coupling, the cavity frequency, the density and the density gradient.

When interpreting these results, it is important to consider the properties of the actual cavity setup, which, given its finite finesse, can couple only a finite volume of matter coherently~\cite{svendsen2023theoryquantumlightmatterinteraction}. This maximum volume cuts off the notion of an infinite system size thermodynamic limit, and the effect of the light-matter correlations presented here should always be put into relation with a finite system smaller than that volume. For a system larger than the maximum volume, the finite coherence length of the cavity photons has to be taken into account. For very short coherence lengths of the order of a few unit cells, the supercell extrapolation in the present work would also be cut off. In such cases, a full minimal-coupling treatment is likely necessary.
The cavity properties also determine the realistic values for $x_\epsilon$. To resolve the crossover to the $x_\epsilon \to \infty$ limit, we simulated parameter sets deep into the diamagnetic regime $\tilde{\Omega}/\Omega \gg  1$, whereas realistic values for $x_\epsilon$ for many materials and cavities may be much smaller. More detailed investigations of the small $x_\epsilon $ regime are left for future work.

Apart from the energy functional itself, we find that for the given Hamiltonian, the weak-coupling perturbation theory at constant renormalized frequency $\tilde{\Omega}$, as opposed to the perturbation theory that uses the bare frequency, is remarkably effective at describing our data. 
This provides a key many-body benchmark for future studies.
We expect that existing perturbative functionals, in particular those in the dipole gauge, can be improved by treating $\tilde{\Omega}$ as a constant during the expansion. Further, we find that the infinite-coupling limit is a well-defined model with a finite light-matter correlation energy, albeit this limit is approached rather slowly.

While this energy functional, which is derived from a simplified model,
is only the first step towards a 
general
functional that is
applicable for all ab-initio QED material calculations, it should
already be useful for quasi-2D systems in combination with regular electronic structure functionals. The extensions to three-dimensions, multiple modes and Coulomb interactions are in principle all within reach 
with
our AFQMC method,
and are left for future works.
Beyond the implications for QEDFT, our work also highlights the viability of QED-AFQMC itself as a tool for simulating cavity-coupled bulk systems, which is of particular interest for scenarios where both electron-photon and electron-electron correlations are significant.

\textit{Acknowledgements}.
We thank Michael Ruggenthaler, Mark Kamper Svendsen, Martin Claassen, Christian Eckhardt, Daniele Guerci, and I-Te Lu for fruitful discussions. The Monte Carlo simulations are based on the Carlo.jl\cite{weber_carlo_2024} framework. We used Optim.jl~\cite{mogensen_optim_2018} and Zygote.jl~\cite{Zygote.jl-2018} for optimizing our trial wave functions and Makie.jl~\cite{danisch_makie_2021} for creating the figures. L.W. acknowledges support by the Deutsche Forschungsgemeinschaft (DFG, German Research Foundation) through grant WE 7176-1-1. The Flatiron Institute is a division of the Simons Foundation.

 \bibliography{paper.bib}

\begin{center}
\textbf{\large Supplemental Material}
\end{center}

\setcounter{equation}{0}
\setcounter{figure}{0}
\setcounter{table}{0}
\setcounter{page}{1}
\section{Perturbative regimes}
In this section, we will provide a short derivation for (i) the perturbative expression of the light-matter correlation energy in the weak-coupling limit and (ii) the effective Hamiltonian in the infinite coupling limit. For both cases, it is best to start out from the modified Hamiltonian
\begin{align}
\label{eq:tildehamiltonian}
H &= \frac{1}{2} \sum_i \mathbf (\mathbf p_i^2 +\Vext(\mathbf r_i))  + \frac{\Qph \pol}{\sqrt{V_\text{c}\tilde\Omega}} \cdot \Pt \nonumber\\
&+ \frac{\tilde\Omega}{2} (\Piph^2 + \Qph^2) - \frac \Omega 2.
\end{align}
after a canonical transformation that absorbs the $\mathbf{A}^2$ term into a renormalized frequency $\tilde\Omega = \Omega \sqrt{1+N\pol^2/V_\text{c}\Omega^2}$
\subsection{Weak coupling regime}
In the limit of $x_\epsilon\ll 1$ at a constant $\tilde\Omega$, we can perform a perturbative expansion, similar to the one that has been considered in \cite{lu_electronphoton_2024}.

At zeroth order, the eigenspectrum is given by the product states $\ket{\Psi_{\{n\};k}} = \ket{\psi_{n_1}\cdots\psi_{n_N}}\otimes \ket{k}$ that consist of electronic Slater determinants and $k$-photon states. We call their corresponding energies $E_{\{n\},k}$ and assume that $n$ labels the single-particle eigenstates in increasing order and $\{n\}$ denotes an $N$-set of them.

We now want to compute the change in the ground state energy $E$ of the coupled system. The linear order in $|\pol|$ of this correction vanishes---even for finite-current states on a torus.

Thus, in leading order we have
\begin{align}
E^{(2)} &= - \frac{1}{V_\text{c}\tilde\Omega} \!\!\!\!\sum_{\substack{\{n\};k\\\neq \{1..N\};0}}\!\!\!\!\frac{|\braket{\Psi_{\{n\};k}|\Qph \pol \cdot \Pt| \Psi_{\{1..N\}; 0}}|^2}{E_{\{n\};k} - E_{\{1..N\};0}}.
\end{align}
Here, only $k=1$ leads to finite contributions. However in the electronic sector, we need to distinguish the cases $\{n\} = \{1..N\}$ and $\{n\} = \{1,\dots,h-1, h+1,\dots, N,p\}$, corresponding to the particle-hole excitation from $h\le N$ to $p>N$.
\begin{align}
\label{eq:secondorderpert}
E^{(2)} = &- \frac{1}{2V_\text{c}\tilde\Omega^2} \qty(\sum_{i=1}^N \braket{\psi_i|\pol\cdot\mathbf{p}|\psi_i})^2\nonumber\\
	&-\frac{1}{2V_\text{c}\tilde\Omega} \sum_{\substack{h\le N\\p>N}} \frac{|\braket{\psi_p|\pol \cdot \mathbf{p}|\psi_h}|^2}{\tilde\Omega + \varepsilon_p - \varepsilon_h}.
\end{align}
The first term can be related to the perturbative correction to the energy of a light-matter product state, $\ket{\Psi_\text{CS}} = \ket{\psi_\text{el}}\otimes\int \frac{dq}{\sqrt[4]{\pi}} e^{-(q-q_0)^2/2}\ket{q}$,
\begin{equation}
E_\text{CS} = \braket{\Psi_\text{el}|\sum_{i=1}^N \frac{\mathbf p_i^2}{2} + \Vext(\mathbf r_i) + \frac{q_0 \pol\cdot \mathbf p_i}{\sqrt{V_\text{c} \tilde\Omega}}|\Psi_\text{el}} + \frac{\tilde\Omega}2q_0^2,
\end{equation}
where $q_0$ is the coherent displacement (unlike in the main text, $\ket{\Psi_\text{CS}}$ is not squeezed because we have already absorbed the $\mathbf{A}^2$ term). The energy is minimized by $q_0 = -\braket{\pol \cdot \Pt}/\sqrt{V_\text{c}{\tilde\Omega}^3}$, yielding
\begin{equation}
\label{eq:ECSfunct}
E_\text{CS} = \braket{\Psi_\text{el}|\sum_{i=1}^N  \frac{\mathbf p_i^2}{2} + \Vext(\mathbf r_i) |\Psi_\text{el}} - \frac{\braket{\Psi_\text{el}|\pol\cdot \mathbf p_i|\Psi_\text{el}}^2}{2V_\text{c}{\tilde{\Omega}}^2}.
\end{equation}
To leading order we have $\ket{\Psi_\text{el}} = \ket{\Psi_{\{1..N\}}}$, which matches exactly with the first term of \cref{eq:secondorderpert}, so that the perturbative correlation energy, $\Ec = E - E_\text{CS}$, is the one given in the main text.
\subsection{Strong coupling}
In the strong coupling regime, a perturbative expansion is less straight forward, because of the interplay of the limit $|\pol|\to\infty$ with the unbounded operators in \cref{eq:tildehamiltonian}. This issue can be circumvented by first performing the Lang-Firsov transformation,
\begin{equation}
U=\exp(i\Piph \pol\cdot\Pt/\sqrt{V_\text{c} \tilde{\Omega}^3})
\end{equation}
that cancels the light-matter interaction term.
\begin{align}
H_\text{LF} &= U^\dag H U \nonumber\\
&= \frac{1}{2} \sum_i \qty[ \mathbf p_i^2 +\Vext\qty(\mathbf r_i + \frac{\Piph \pol}{\sqrt{V_\text{c}\tilde\Omega^3}})] - \frac{1}{2V_\text{c}\tilde\Omega^2} (\pol\cdot\Pt)^2 \nonumber\\
&+ \frac{\tilde\Omega}{2} (\Piph^2 + \Qph^2) - \frac \Omega 2.
\end{align}
Noting that $\tilde\Omega \sim \sqrt{N/V_\text{c}}|\pol|$ as $|\pol|\to\infty$, we see that the shift of the argument of $\Vext$ (dubbed mollification of the potential in Ref. \cite{schafer_making_2021}), vanishes in the strong coupling limit. Introducing a normalized polarization vector $\hat{\pol} = \pol/|\pol|$, we can then write the infinite coupling Hamiltonian.
\begin{align}
H_\infty &= \lim_{|\pol|\to\infty} H_\text{LF}\nonumber\\
&= \frac{1}{2} \sum_i \qty[ \mathbf p_i^2 +\Vext\qty(\mathbf r_i)] - \frac{(\hat{\pol}\cdot\Pt)^2}{2N},
\end{align}
where we omit the (infinite) shift of the vacuum energy that will, in any case, drop out in the calculation of the light-matter correlation energy.

One might wonder if the infinite coupling ground state actually exists. However, for $\Vext=0$, it is straightforward to see that $H_\infty$ is still bounded from below. Since $\Vext(\vb r_i)$ itself is bounded, so is the complete Hamiltonian. We also see that compared to a usual electron-electron interaction, due to the single-mode approximation, the interaction of $H_\infty$ only couples electrons with the same momentum, making it a factor $N$ weaker than e.g. the Coulomb interaction that couples a macroscopic number of momentum transfers. This explains why even in the infinite coupling case, the light-matter correlation energy remains intrinsic. 

To derive an expression for $\Ec$, we further need to derive the infinite coupling limit of $E_\text{CS}$: by taking the infinite coupling limit in \cref{eq:ECSfunct}, we can see that this energy corresponds exactly to the Hartree-Fock energy of $H_\infty$. 
\section{Asymptotic behavior of $\mathcal{Q}^2$}
In this section, we discuss the asymptotic behavior of the quantity $\mathcal{Q}^2$ for the case of large $v$ in our model to explain the fitting function  $x_q(x_v)$ we used in the main text.

In the limit of large $v$, each electron becomes localized in one of the minima of $\Vext$, which we can expand as $\Vext \approx 2\pi^2 v \vb{r}^2/a^2$. Consequently, the density $n(r)$ will be given by that of the ground state of the harmonic oscillator,
\begin{equation}
    n(\vb r) = \sqrt{\frac{2\sqrt{v}}{a}} e^{- \frac{2\pi\sqrt{v}\vb{r}^2}{a}}
\end{equation}
from which follows that $\mathcal{Q}^2 \sim \sqrt{v}/a$.

We further know that $\mathcal{Q}^2(v) = -\mathcal{Q}^2(-v)$ since the two cases are connected by translation symmetry. Assuming that the point $\mathcal{Q}^2(0)=0$ is not singular, for small $v$, we expect $\mathcal{Q}^2\sim v^2$.

The fitting function $x_q(x_v)$ fulfills both of these asymptotics.
\end{document}